\documentclass[preprint,12pt]{elsarticle}
\journal{Physics of the Dark Universe}
\usepackage{graphicx}
\usepackage{xcolor}

\begin{document}

\begin{frontmatter}

\title{Hawking-Radiation Recoil of Microscopic Black Holes}


\author[1,2]{Samuel Kov\'a\v{c}ik\corref{cor1}%
}
\ead{samuel.kovacik@fmph.uniba.sk}

\address[1]{Faculty of Mathematics, Physics and Informatics, Comenius University in Bratislava, Bratislava, Slovakia}
\address[2]{Department of Theoretical Physics and Astrophysics, Faculty of Science, Masaryk University, Brno, Czech Republic}

\begin{abstract}
Hawking radiation would make microscopic black holes evaporate rapidly, which excludes them from many astrophysical considerations. However, it has been argued that the quantum nature of space would alter this behaviour: the temperature of a Planck-size black hole vanishes, and what is left behind is a Planck-mass remnant with the cross-section of $\approx 10^{-70}\mbox{m}^2$, which makes a direct observation nearly impossible. Such black hole remnants have been identified as possible dark matter candidates. Here, we argue that the final stage of evaporation has a recoil effect that gives them a velocity up to $\approx 10^{-1} c$. This would lead to a disagreement with the cold dark matter cosmological model unless the primordial black hole formation ceased shortly after the inflation era.
\end{abstract}


\end{frontmatter}

\section{Introduction}
Even though the quantum theory of gravity has not yet been established, we can make assumptions about some of its properties. One of them is the assumption that space has a structure that becomes evident at the Planck scale. This idea is not new and has been explored from various angles \cite{Connes:1996gi, Doplicher:1994tu, Groenewold:1946kp, Madore, Moyal:1949sk, Presnajder, Snyder:1946qz}.

A general feature of theories of quantum space is the impossibility of distinguishing points with a separation on the order of the Planck length. The photon or another particle used to perform the measurement would be hidden under its own Schwarzschild radius and would form a microscopic black hole, preventing the extraction of any information. Infinitely short distances and infinitely large energies are expected to be absent from the theory.

In physics, we often rely on the notion of point-particle matter density (the Dirac delta function), often as a mathematical tool or as an approximation. Yet, there is one prominent example where it is taken to describe the physical reality: the matter distribution of a black hole is zero everywhere but at one point. Theories of quantum space often lack the notion of exact point localisation and any matter distribution is rendered nonsingular; regular black holes have been studied before, for example in \cite{Dymnikova:1992ux}. This modifies the behaviour of black holes of a size that is comparable to the fundamental length scale. One of the striking differences compared to the ordinary black hole theory is that the Hawking temperature, defined to be proportional to the surface gravity at the horizon, does not grow indefinitely but instead drops to zero at small but positive mass, resulting in a microscopic black hole remnant \cite{Carr:2009jm}.

Black holes remnants have been considered as possible dark matter constituents \cite{Chen:2002tu}. In ordinary space, small black holes evaporate rapidly. In a space with a quantum structure, they can be eternal and are very difficult to detect due to their small cross-section. If they contributed significantly to the overall dark matter density, proving it would be difficult as their direct detection seems to be impossible.

A gravitational collapse leading to a black hole formation or a merger of two black holes can produce gravitational waves that carry significant momentum in a single direction, recoiling the resulting black hole the opposite way \cite{Bekenstein:1973zz, Campanelli:2007ew}. It has been recognized that the thermal radiation has a recoil effect that causes a position drift \cite{Page:1979tc, Nomura:2012cx, Flanagan:2021svq,Dai:2007ki, Mack:2019bps}. Here we argue that relativistic recoil velocities are a general feature of black holes with a density profile extended on the scale of the Planck length. To be a plausible cold dark matter candidate, their formation must have ceased shortly after the Big Bang for the expansion of space to be able to slow them down. If we consider the primordial black hole mass spectrum, we can conclude their formation had to cease before $ t \sim 10^{-28}\ \mbox{s}$.

\section{Planck-size black holes}
Two points separated by approximately the Planck length cannot be distinguished as any such effort would result in a production of a microscopic black hole. Therefore, in a similar consideration that led to quantisation of phase space in quantum mechanics, one can suspect a quantised structure of the ordinary space where infinitely short distances and infinite energies are absent.

Both of those vices, infinitely short distances and infinitely large energies, are present in the ordinary description of black holes. Their mass is located in a single zero-size point, where the density has to be infinite. Also, the temperature of the Hawking radiation grows to infinity during the final moment of the evaporation. The quantum structure of space renders the matter distribution regular and prevents the black hole from reaching infinite temperature.

There were various attempts to investigate the consequences of the quantum nature of space on the behaviour of black holes, e.g., \cite{Kovacik:2017tlg, Nicolini:2005vd,Nicolini_2005}. Various models of space modified on the length scale $\lambda$ may provide different forms of matter density, $\rho(r)$, that reproduce the singular delta-function in the $\lambda \rightarrow 0$ limit. We will assume $\lambda$ to be the Planck length and express all other dimensionful quantities in terms of Planck units.

\begin{figure}[h] 
\centering
  \includegraphics[width=1.0\textwidth]{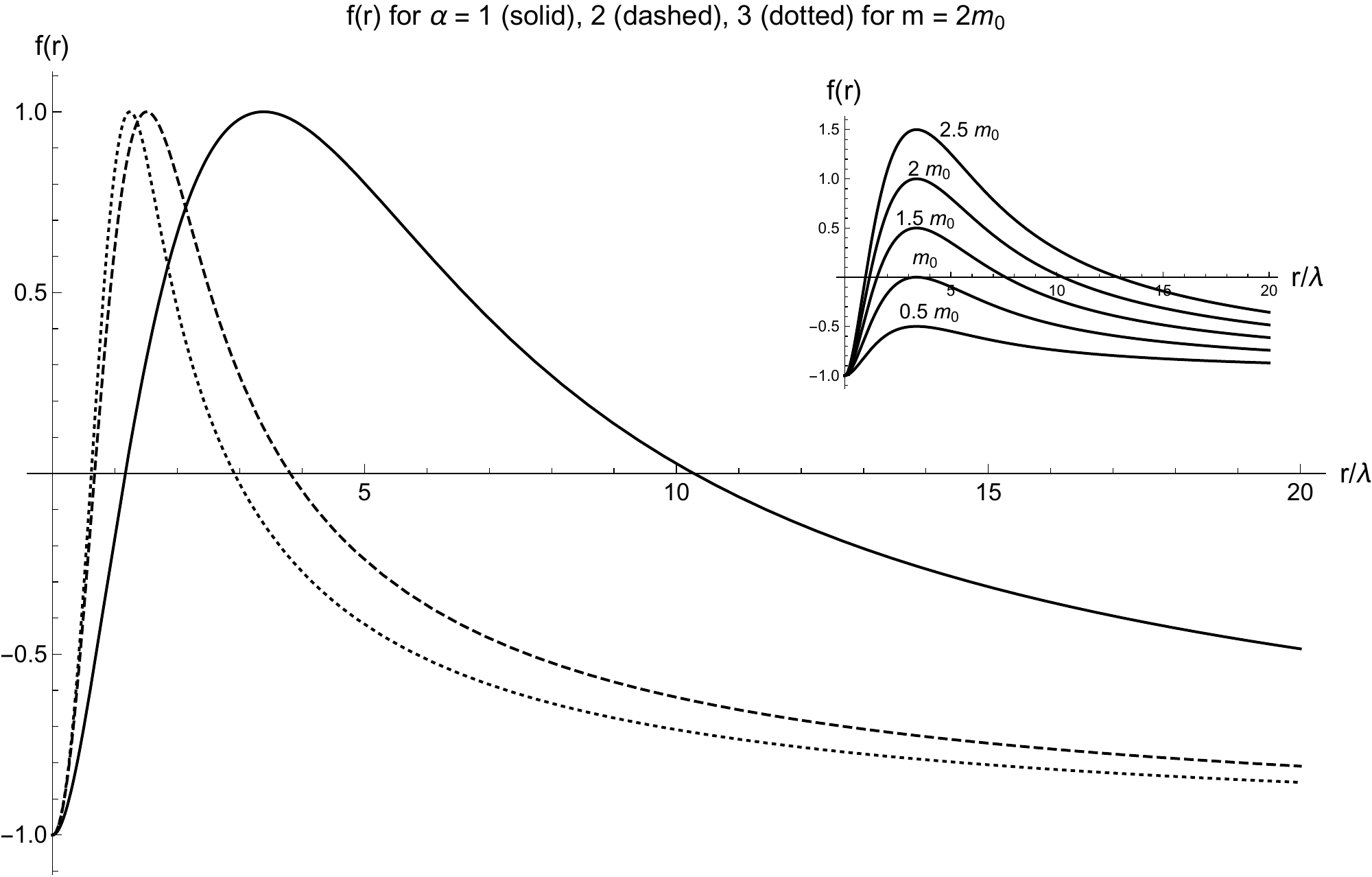}
    \caption{\scriptsize{The function $f(r)$ for the three different choices of matter density $\rho_\alpha(r) = \rho_0 e^{- \left(r/\lambda\right)^\alpha}$ evaluated at $m=m_0$, where $m_0$ is the minimal black hole mass. The zero points $f(r_\pm)=0$ mark the event horizons. For $m\gg m_0$ one of them is close to the origin $r_- \approx 0$ and the other is close to the Schwarzschild value of $r_+ \approx 2m$. As the mass is decreased, they move toward each other and meet for $m=m_0$. At this point, the derivative $f'(r)$ vanishes at the horizon and so does the Hawking temperature. For $m<m_0$, $f(r)=0$ has no solution so there is no horizon. The small plot shows $f(r)$ for $\alpha=1$ and various multiples of the critical mass $m_0$.}}
   \label{figure f}
\end{figure}

One can then take this matter distribution, $\rho(r)$, and insert it into the Einstein field equations, looking for a solution in the Schwarzschild-like form $g=\mbox{diag} \left(f(r),-f^{-1}(r), r^2, r^2 \sin \theta\right)$ (with coordinates $(t,r,\theta,\varphi)$ and signature $(-,+,+,+)$). This approach has been used for matter densities of the form $\rho_\alpha(r) = \rho_0 e^{- \left(r/\lambda\right)^\alpha}$, where the values $\alpha=1,2$ have been derived in the context of noncommutative space \cite{Kovacik:2017tlg, Nicolini:2005vd,Nicolini_2005}, and the $\alpha=3$ case from the consideration of vacuum polarisations \cite{Dymnikova:1992ux}. Recently some other values of $\alpha$ have been considered as the Einasto density profiled derived in the dark matter halo context \cite{Batic_2021}.  The value of $\rho_0$ is fixed in each of those cases separately. The corresponding Einstein field equation is
\begin{equation} \label{diff eq for f}
\frac{1 + f(r) + r f'(r)}{r^2} = 8 \pi \rho(r),
\end{equation}
we are using units such that $G=c=k_B=1$. Solutions that satisfy $f(0)=-1$ and $f(r) \approx -1 + \frac{2m}{r}$ for $r \gg \lambda$ have been found  in the aforementioned references. They share a common property: there is a critical mass $m_0 \sim \lambda$ for which the solution has only one horizon defined by $f(r_0)=0$. For any larger mass $m > m_0$ there are two horizons $r=r_\pm$ for which $f(r_-)=f(r_+)=0$. For any smaller mass $m<m_0$, there is no solution to $f(r)=0$ and therefore no horizon. For masses $m \gg m_0$, the larger horizon is close to the Schwarzschild value $r_+ \approx 2m$ and the smaller is close to the origin $r_- \approx 0$. The function $f(r)$ for the three considered cases is shown in the figure \ref{figure f}, we will discuss solutions to other matter distributions later in this paper. \\
\hspace*{.4cm} The Hawking temperature is defined, in the usual manner, by the surface gravity at the outer event horizon and has been shown before to have an upper boundary for the aforementioned cases. We will explicitly show that this also holds for three infinite classes of regular matter distributions and discuss under what requirements will the Hawking radiation always be finite. 

\begin{figure}[h] 
\centering
  \includegraphics[width=1.0\textwidth]{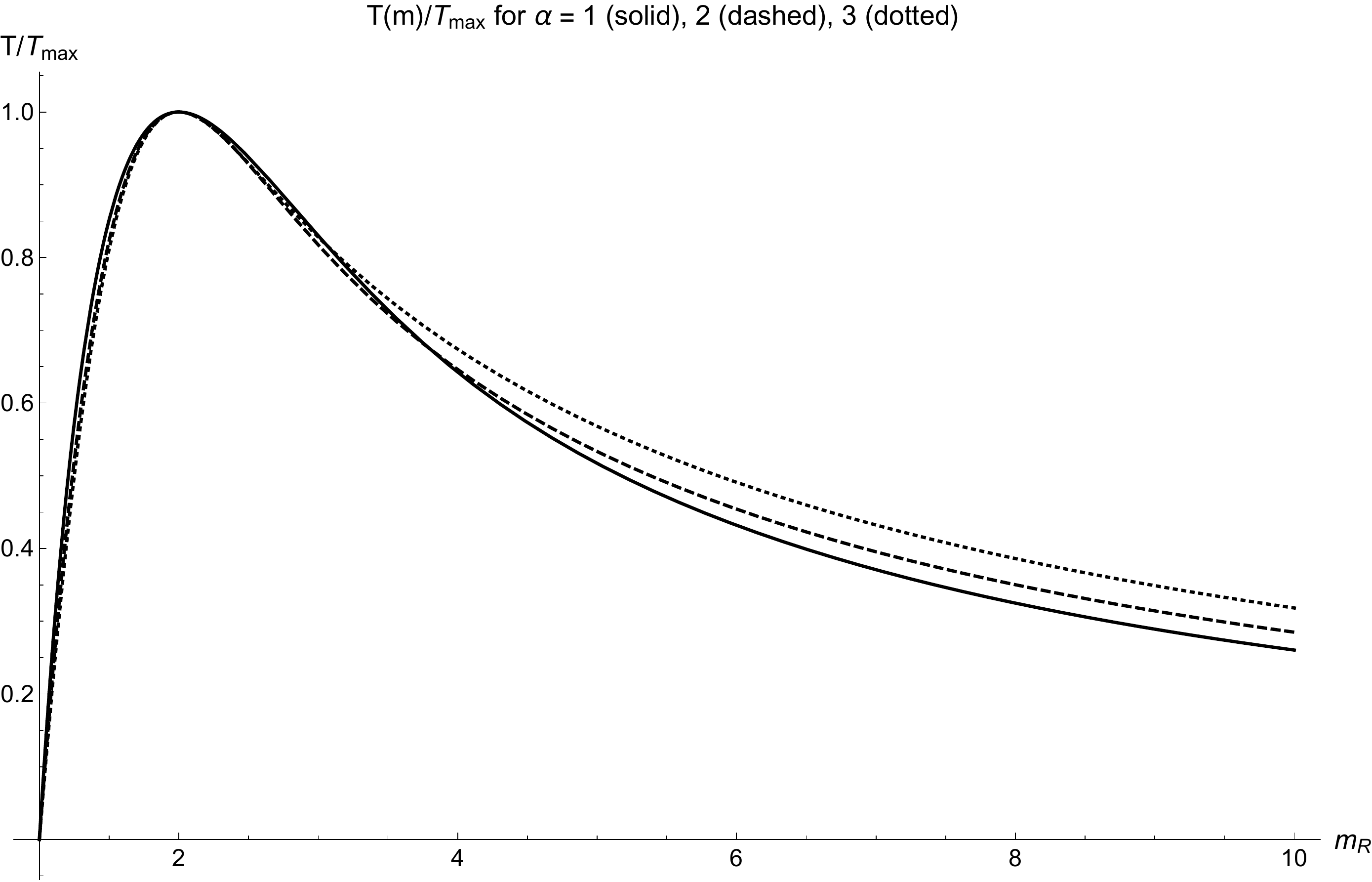}
    \caption{\scriptsize{ 
The Hawking temperature of the black hole with matter density $\rho_\alpha(r) = \rho_0 e^{- \left(r/\lambda\right)^\alpha}$ expressed in terms of the mass, $m_R$, and the maximal temperature, $T_{\mbox{\scriptsize{max}}}$. The mass $m_R$ has been rescaled and shifted so $m_0$ and $m_{\mbox{\tiny{Tmax}}}$ match across different profiles. For the considered cases with $\alpha=1,2,3$ we have $m_0  \stackrel{.}{=} (2.57/0.95/0.73)\lambda$ and $T_{\mbox{\scriptsize{max}}} \stackrel{.}{=} (0.01/0.03/0.04)\lambda^{-1}$, where $\lambda^{-1}$ is the Planck temperature. Note that when expressed in dimensionless units, all three cases have nearly identical temperature profiles. All solutions discussed in this paper, also considering more general matter distributions, have similar temperature profiles.}}
   \label{figure T}
\end{figure}

\section{Hawking radiation recoil}
As expected due to the removal of arbitrary small distances and therefore arbitrary short wavelengths, the infinite temperature is avoided. Instead of growing indefinitely the temperature drops to zero as $m$ approaches the minimal mass $m_0$ and the black hole becomes frozen. In the figure \ref{figure T}, we show the temperature dependence for the three considered cases rescaled with respect to the minimum mass $m_0  \doteq (2.57/0.95/0.73)\lambda$ and the maximal temperature $T_{\mbox{\scriptsize{max}}} \stackrel{.}{=} (0.01/0.03/0.04)\lambda^{-1}$, where $\lambda^{-1}$ is the Planck temperature. 

Another shared feature of the considered models of regular black holes is that the temperature, as a function of mass, grows rapidly in the vicinity of $m_0$. This means that when the black hole reaches the mass $m_{\mbox{\tiny{Tmax}}}$, for which it has the maximal temperature, the radiation is so energetic that the mass difference $\Delta m = m_{\mbox{\tiny{Tmax}}}-m_0$ is radiated in a relatively small amount of quanta, $N_q \approx \frac{\Delta m}{T_{\mbox{\scriptsize{max}}} } \le 10^2$.

Each quantum carries the momentum of $p \approx \frac{\Delta m}{N_q}$ and due to the conservation law the black hole receives the opposite momentum. As the radiation is directed randomly, so are the momentum impulses on the black hole. The black hole performs a random walk in the momentum space, and after radiating $N_q$ quanta it will carry a momentum $p_{\mbox{\small{rec}}} \approx \frac{\Delta m}{\sqrt{N_q}}$. As a result, its final recoil velocity will be 
\begin{equation} \label{rec}
v_{\mbox{\small{rec}}} \approx  \frac{\Delta m}{m_0 \sqrt{N_q}}
\end{equation}
This is the recoil effect due to thermal Hawking radiation of Planck-size black holes. For the considered cases of $\rho_\alpha$, we have $\frac{\Delta m}{m_0} \doteq 0.33/0.26/0.22$ and $N_q \doteq 88/8.4/3.8$; therefore $v_{\mbox{\small{rec}}}\doteq 0.035/0.091/0.112\ c$. In all three cases, this is a considerable fraction of the speed of light. 

Unless some slowing-down process took place after the formation of microscopic black holes, such as a considerable expansion of space, we have to expect them to have significant velocities. This can have implications on how they fit the role of the cold dark matter constituents.

\section{Cosmological consequences}
The recoil effect due to Hawking radiation modified by the quantum structure of space discussed here could make the Planck-size black holes improbable dark matter candidates as they would obtain velocities large enough to be incompatible with the properties of cold dark matter favoured by observations. However, it was pointed out by the authors of \cite{Lehmann:2021ijf} that the velocity would be reduced by the expansion of space, therefore this effect does not exclude black holes that evaporated shortly after the Big bang when the scale factor was $a_F < 10^{-7}$. Also, if the space expansion would not slow them down, they would exceed escape velocities from most astronomical objects and would be gravitationally unbound. 

There are various models of production and behavior of black holes during the earliest stages of the Universe \cite{Green:2020jor, Konoplich:1999qq,Dymnikova:2015yma, Clesse:2017bsw, Hawking:1974rv}. As we have mentioned, high recoil velocity excludes models in which the evaporation is finished later. The commonly considered mechanism of the primordial black formation is the collapse of overdensities during the radiation era that resulted from primordial fluctuations from the inflation era. There are also other options, for example the cosmic string collapse \cite{Caldwell:1995fu}, the bubble nucleation \cite{Kusenko:2020pcg} or the domain wall collapse \cite{Khlopov:2008qy}. 

The aforementioned bound, $a_F < 10^{-7}$, translates into the final moment of evaporation $t_F \approx 10^{6}  \mbox{s}$. Black holes able to evaporate in such a short time must have been created with the mass $m_{PBH} \approx 10^{10}\mbox{g}$ at most. The classical result from \cite{Carr:1975qj} states that black holes formed at the time $t$ during the radiation era have $m_{PBH} \approx 10^{38} \ t  \left[\frac{\mbox{g}}{\mbox{s}}\right]$. That means that the last possible moment of formation of black holes satisfying the bound is $t \approx 10^{-28} \mbox{s}$. This suggests that the formation of  small primordial black holes had to be completed shortly after the inflation era. 

\section{Generalisation of the results}
So far we have discussed the recoil effect only for the three different matter densities that previously appeared in the literature. We will now investigate large classes of regular matter distributions to show that our result is not dependent on that particular choice. 

The starting point is a regular density that is normalised as 
\begin{equation} \label{norm}
\int \limits_0^\infty \tilde{\rho}(r) 4 \pi r^2 dr = 1.
\end{equation}
Separating the mass from the matter density, $\rho(r) = m \tilde{\rho}(r)$, will prove beneficial shortly. We will consider three different classes
\begin{eqnarray} \label{dens}\nonumber
\tilde{\rho}_1(r; q) &=& n_1 e^{-\left(r/\lambda\right)^q},\\
\tilde{\rho}_2(r; q) &=& n_2 \left(1+\left(r/\lambda\right)\right)^{-q},\\ \nonumber
\tilde{\rho}_3(r; q) &=& n_3 \left(1+\left(r/\lambda \right)^q\right)^{-1},
\end{eqnarray}
with the range of $q$ for which the normalisation (\ref{norm}) is possible. The regular solution to (\ref{diff eq for f}) is then of the form
\begin{equation} \label{f int}
f(r) = -1 + \frac{2m}{r}\int \limits_0^r \tilde{\rho}(r') 4 \pi r'^2 dr'.
\end{equation}
This makes it evident that $f(0)=-1$ and that for $r \rightarrow \infty$ the solution asymptotes to the Schwarzschild one. Using this form it is also easier to separate classical and quantum effects caused by having $\lambda > 0$. The three considered matter distributions lead to solutions $f_i(r;m,q) = -1 + \frac{2m}{r} g_i(r;q)$ with
\begin{eqnarray} \nonumber
g_1(r;q) &=& \frac{3\ \left( \Gamma\left(3\ q^{-1}\right) - \Gamma\left(3\ q^{-1},r^q\right)\right)}{q \ \Gamma\left(1 + 3\ q^{-1}\right)} ,\\  
g_2(r;q) &=&1 - \frac{ \left(2 + \left( q-1\right) r \left(2 + \left(q-2\right) r\right)\right)}{2\left(1 + r\right)^{q-1}},\\ \nonumber
g_3(r;q) &=& \frac{q\ r^3 \ \setlength\arraycolsep{1pt}
{}_2 F_1 \left(1,3\ q^{-1},1+3\ q^{-1},-r^q\right) \sin\left(3 \pi\ q^{-1}\right) }{3\pi}  ,
\end{eqnarray}
where $\Gamma(a,b)$ is the upper incomplete Gamma function and $\setlength\arraycolsep{1pt}
{}_2 F_1$ is the hypergeometric function. 

This form of the solution makes it easy to read off the behaviour, details of the functions $g_i(r;q)$ are not important. One just needs to note that the combination $r^{-1}g(r;q)$ is $0$ for both $r\rightarrow 0$ and $r \rightarrow \infty$. Somewhere in between it reaches a single maximum and the value of $m$ determines whether at this maximum is the function $f_i$ positive (two horizons), zero (single horizon) or negative (no horizon). Also, the value of $m$ determines the steepness of $f_i$ at the outer horizon and therefore also the Hawking temperature. 

How many maxima of $f(r)$ can there be? For the considered class of solutions there was a single one but we can show that the same holds for many other choices of $\rho (r)$. We will assume that $\rho (r)$ is a decreasing function. Let us denote $f(r) = - 1 + 2m y(r)$ where $y(r) = r^{-1} h(r)$ and $h(r)$ is the integral from the equation (\ref{f int}). At extremum we have $y'(r)=0$ and $y''(r) = r^{-1} h''(r) = 4 \pi \left( 2 \rho(r) + r \rho'(r)\right)$ determines whether it is a maximum, a minimum or an inflection point.

For there to be more than one maximum, the combination $j(r) =2 \rho(r) + r \rho'(r)$ would need to have at least two roots as it would change sign at least twice. In the considered cases we have $j_\alpha(r)$ with a single root $r_I$    
\begin{eqnarray} \nonumber
j_1 (r) \stackrel{\sim}{=} 2 - q \left( r/\lambda \right)^q &\rightarrow & r_{I,1} = \lambda \left( 2 q^{-1} \right)^{q^{-1}}, \\ 
j_2 (r) \stackrel{\sim}{=}  \left( 2 \left(1 + \left( r/\lambda \right)  \right) - q \left( r/\lambda \right) \right) &\rightarrow & r_{I,2} = \lambda\left( \frac{2}{q-2}\right), \\ \nonumber
j_3 (r) \stackrel{\sim}{=} 2\left(1 + \left( r/\lambda \right) ^q \right) - q \left( r/\lambda \right) ^q & \rightarrow & r_{I,3} = \lambda \left( \frac{2}{q-2}\right)^{q^{-1}},
\end{eqnarray}
where $\stackrel{\sim}{=}$ means that we have removed parts of $j(r)$ unimportant for the determination of the roots. This shows why many properties of regular black holes seem to be common. However, the generality of some features, such as the sharp temperature drop at the end of the evaporation process, is more intricate.      

\begin{figure}[h] 
\centering
  \includegraphics[width=1.0\textwidth]{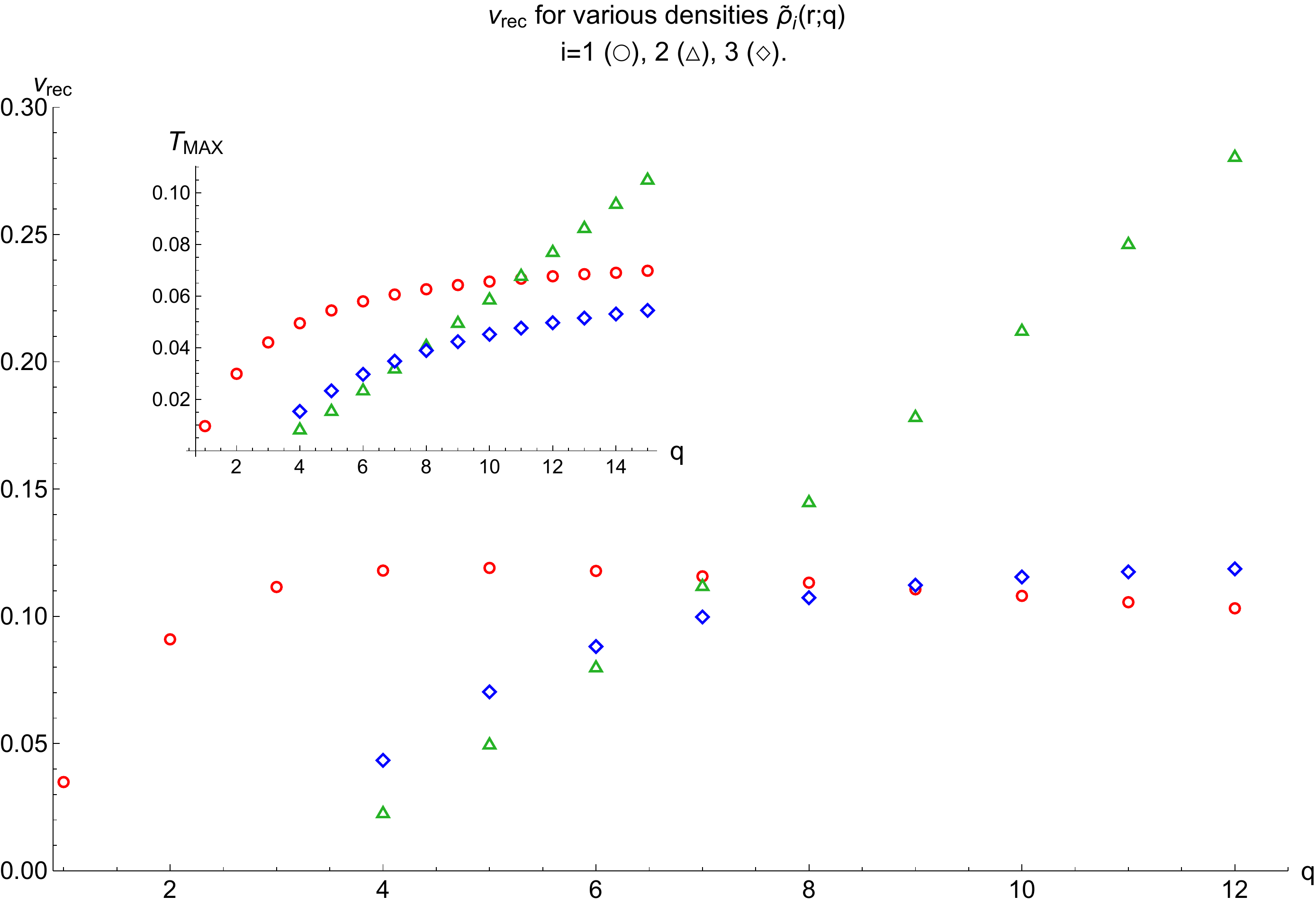}
    \caption{\scriptsize{The recoil velocity for different densities from (\ref{dens}) and values of $q$ up to $12$. Values of $v_{\mbox{\scriptsize{rec}}}$ range from $\approx 0.02c$ to $\approx 0.28 c$. In the small plot we show the maximal temperature reached in those cases.}}
   \label{figure vr}
\end{figure}

\section{Conclusion}
The are three main results of this paper. The first is the derivation of the recoil effect due to the Hawking radiation for microscopic black holes which is caused by a nontrivial structure of the space on the scale of the Planck length. The second is setting the constraint on the last possible moment of the primordial black hole formation to be shortly after the end of the inflation era. The third main result is finding solutions for various regular matter densities and establishing the generality of the previous results. 

We have focused on the simplest choices of matter density distributions, some of which have been already studied in the literature. Using (\ref{f int}), one can construct a solution for any regular and normalizable matter density $\rho$ that could arise in a physical context. We expect that the behaviour for other choices of monotonic functions $\rho$ would be similar to what we have observed, but the situation can be different for cases of nonmonotonic functions, for example $\rho(r;a,b) \sim \frac{(r/\lambda)^a}{1+(r/\lambda)^b}$.

The densities that we have studied, $\rho_i$, were parametrised by the length scale $\lambda$, the mass $m$, and the parameter $q$. In the $q\rightarrow \infty$ limit, the distribution $\rho_2$ becomes singular. The other two choices lead, in this limit, to a uniform distribution within a sphere $r < \lambda$ and could be relevant in scenarios where the space has a foam-like structure and the black hole mass can uniformly occupy a single bubble.

This paper only considers Planck-size black holes, the expected behaviour of larger black holes, such as those analysed in \cite{DeLuca:2020agl}, is not affected by the present discussion. Our results seem to be rather general and should hold up to a small numerical factor as we neglected some details, for example, the exact three-dimensional random-walk formula or the contribution to the recoil effect from the stages of $T>T_{\mbox{\scriptsize{max}}}$. From the classical work \cite{Page:1976df} we know that Hawking radiation has a graybody factor \cite{Boonserm:2014rma}, which is however hard to estimate as it can be sensitive to unknowns physics such as the existence of extra spatial dimensions \cite{Hod:2011zzb} or the dark particle sector \cite{Bernal:2020kse}. Our discussion has not been very detailed as the exact theory of quantum gravity is not known and neither is the behaviour of the Hawking radiation on its scale. However, at least under the current assumptions, the recoil effect due to the thermal radiation of microscopic black holes should be taken into consideration as it seems to be a general feature of regular microscopic black holes. 

\section*{Acknowledgement}
This research was supported by VEGA 1/0703/20 and the MUNI Award for Science and Humanities funded by the Grant Agency of Masaryk University. Valuable comments from V. Balek, J. Tekel and N. Werner were greatly appreciated.

\bibliographystyle{unsrt}
\bibliography{References}

\end{document}